%% file: main.tex
\documentclass[14pt]{article}
\usepackage[utf8]{inputenc}
\usepackage{xcolor}
\usepackage{hyperref}
\usepackage{graphicx}
\usepackage{amsmath}
\usepackage{amssymb}
\usepackage{algorithm}
\usepackage{algorithmic}
\usepackage{array}
\hypersetup{
    colorlinks=true,
    linkcolor=blue,
    filecolor=magenta,      
    urlcolor=cyan,
    pdftitle={Overleaf Example},
    pdfpagemode=FullScreen,
    }
    
\usepackage[left=2.15cm, right=2.15cm, bottom=2.15cm, top=2.15cm]{geometry}

\newcommand{\comment}[1]{{\color{black!50!green} #1}}
\newenvironment {acknowledgements} {  \begin {abstract} } { \end {abstract} }

\title{Lakat: An open and permissionless architecture \\ for continuous integration academic publishing}
\author{Leonhard Horstmeyer \\
Basic Research Community in Physics (BRCP) e.V.}
\date{\today}

\begin{document}

\maketitle

\input{abstract.tex}


\tableofcontents


\input{introduction.tex}

\input{datastructure.tex}

\input{contributions.tex}

\input{protocol.tex}

\input{Integrationadaptability.tex}

\input{conclusion.tex}

\vspace{2ex}
\input{acknowledgements.tex}

'

'

\end{document}

%% file: abstract.tex
\begin{abstract}
    
    In this paper, we present three contributions to the field of academic publishing. Firstly, we introduce Lakat, a novel base layer for a publishing system that fosters collaboration, pluralism and permissionless participation. Drawing inspiration from the philosophy of Imre Lakatos, Lakat is designed as a peer-to-peer process- and conflict-oriented system that supports continuous integration across multiple branches. This architecture provides a robust foundation for the  integration of existing reputation systems and incentive structures or the development of new ones. Secondly, we propose a new consensus mechanism, called Proof of Review, which ensures the integrity and quality of the content while promoting active participation from the community. Lastly, we present Lignification, a new finality gadget specifically designed for branched, permissionless systems. Lignification provides a deterministic way to find the consensual state in these systems, ensuring the system's robustness and reliability in handling complex scenarios where multiple contributors may be proposing changes simultaneously. Together, these contributions aim to provide a convenient starting point to tackle some of the issues in traditional paper-formatted publishing of research output. By prioritizing collaboration, process-orientation, and pluralism, Lakat aims to improve the way research is conducted and disseminated and ultimately hopes to contribute to a healthier and more productive academic culture.

\end{abstract}

%% file: introduction.tex
\section{Introduction}
\label{sc:introduction}

With the vast amount of data structures, of query and storage systems, of versioning and networking tools and of large language models, one may engineer publishing systems by posing certain requirements that give rise to a different and arguably more collaborative, efficient and healthy academic culture. This approach can be contrasted with an incremental adjustment of the existing system, which in many quantitative sciences is called the greedy approach. We propose an architecture leveraging the available technology that we call \textit{Lakat}.
Lakat is a distributed database with a local peer-review consensus layer. The system serves as a permissionless continuous integration solution for collaborative research. One may conceptually think of Lakat as a peer-to-peer version of Wikipedia with a branch structure similar to git and a peer review system.
Our starting point is a set of eight core requirements, that we posit for a publishing system:\\

\indent \textbf{1. Open} -- 
 Content and code base\footnote{Here we refer to the code base of any client implementation.} should be accessible freely\footnote{Internet service providers are not free. So we refer here to additional charges.}.\\
\indent\textbf{2. Permissionless} --
 No one should be barred from contributing.\\
\indent\textbf{3. Pluralistic} -- No monopoly on research opinion.\footnote{This is not not necessarily the same as ``No single source of truth''.}\\
\indent\textbf{4. Process-oriented} -- Emphasizing the process rather than an outcome.\\
\indent\textbf{5. Conflict-oriented} -- Making conflicts a feature rather than a bug.\\
\indent\textbf{6. Curatable} -- Making the presentation and organization of the content part of the output process.\\
\indent\textbf{7. Sustainable} -- 
 Data and compute resources should be low and reuse of fragments encouraged.\\
\indent\textbf{8. AI friendly} -- Allowing all kind of entities to contribute, individuals, groups or AI agents.\\

The research paper, as the gold standard of publicizing research output, poses several threats to the overall scientific endeavor. It is a relic from the times where the printing press had been the latest innovation and where the channels for communicating had a large latency.  We mention six issues associated with the paper-formatted research output that are addressed by Lakat:
\begin{itemize}

 \item It incentivizes the creators of scientific output to withhold preliminary results or results that are either not significant or at odds with a hypothesis. Even if there are significant results\footnote{These may be perceived as significant or later recognized as significant by the community}, they may not meet the eye or mind of other creators or consumers of scientific output until the entire paper has been published. It may then even take on the order of tens of months for the paper-formatted research output to be accessible, which is particularly problematic for impactful research. Thus the process of building on top of previous work and of critical engagement is hindered and in the best case deferred.
 
\item It incentivizes creators to wrap minor changes into the guise of an entire research paper, reusing a possibly templated introduction over and over again. 

\item The output is but a polished snapshot of a process, an inorganic blob "data structure". The process of reaching a result or of not reaching it as well as the review process are generally not part of the output and not naturally representable in the rigid paper-format. The process often doesn't stop with the paper-publication, but continues thereafter and it requires awkward hacks in the form of addenda, corrections or new paper-formatted versions to account for changes.

\item It creates rigid and isolated islands of content, disregarding potentially conflicting or agreeing intersections. Papers address these intersections with citations that are often placed in an unspecific context, and tend to reference an entire paper or body of work rather than a particular part. These intersections between different scientific outputs are not only constrained to citations, but entire paragraphs such as introduction or method sections are often simply replicated from previous papers. Thus, making conflicting or agreeing intersections a manifest part of the data structure can overcome the hacky fixes and shortcomings of the paper-format.

\item The question of who contributed how much to a research output often causes conflict among researchers. A process-oriented publication system facilitates the tracking of contributions and may reduce the cases of unjust allocation of contributorship. In paper-formatted publications the contributorship is proxied by a negotiated ordered list of co-authors, which cannot capture contributions and inevitably leads to unjust allocations. 

\item The effective barring of potential contributors in paper-formatted research does not increase the level of scrutiny, creativity, or quality of the output. On the contrary, maybe another set of eyes can add insights or expand on the results. Why should the self-declared co-authors be in the best position to conduct the research? The fear for the theft of ideas is mostly inherent to bulk-publications and less to process-based research output. 
\end{itemize}

Apart from the abovementioned problems with paper-formatted research, Lakat may also be instrumental for solving other problems with scientific publishing such as the exploitation of scientists regarding their review services and production of output. Even though Lakat does not directly address this, it does provide a base layer upon which a system of incentives can be built.

\subsection{Related Work}

Various solutions have been proposed to improve the process of science publishing with respect to transparency, review, ownership, decentralization, collaboration, openness, and fairness. We exhibit proposed solutions and their benefits or shortcomings. Since Lakat sits at the intersection of branchable version control (c.f. Git \cite{chacon2014pro,git}), large collaborative encyclopedias (c.f. wikipedia \cite{wikipedia}) and peer-to-peer (c.f. human society \cite{decentralizedhumansocieties}) protocols (c.f. Urbit \cite{urbit} or file sharing protocols \cite{distributed_file_sharing, guidi2021data}), we will focus on solutions in that general triangle.

The platform Scholarpedia was launched in 2006 \cite{izhikevich_scholarpedia}. It is a wiki-based format with a peer review layer, where institutional affiliation is required for contribution. It is thus integrating a scholarly component into wikipedia. The requirement of affiliation is also one of the drawbacks of this solution, as it bars some potential contributors. Furthermore, the authors of an article are either chosen or elected. This to our mind has two further problems. First, it raises the question who elects those that elect. Second, the collaborative dimension of wikipedia is lost. In contrast Lakat -- like wikipedia -- retains the permissionless so that no one is barred from editing or from proposing pull requests to change content (see Section \ref{sc:protocol} for details). In 2007 the Citizendium fork of the English wikipedia launched \cite{sanger_citizendium} with the objective to add a quality assurance layer on top of wikipedia. The concept of approved articles played an important role. However, who approves the articles? What happens to subsequent changes? Would they have to be approved again or does the approval yield a sort of finality for the manuscript?
Another wiki-formatted solution is the Manubot platform \cite{himmelstein2019open}, which allows for the collaborative preparation of research articles that can then be sent to peer-reviewed journals. However, Manubot is not a publication platform itself but aids the collaborative process of reaching a traditional publication. 

There are also many attempts to put part of the existing publishing logic onto a cryptographically secure distributed ledger. Everipedia\cite{forselius_everipedia} was a fork of wikipedia. They have also tried to build a quality assurance system on top of it using reputation tokens that can be staked and potentially lost in the process of edits, thus leveraging distributed ledger technology. So instead of tokenizing ownership of edits, they tokenized reputation. Those tokens were deployed on a blockchain (EOS and later Polygon). The project has been archived. Orvium \cite{romero_orvium} on the other hand aims to put submission of manuscripts, revisions and publications onto a blockchain or at least have them stored using some decentralized storage provider. Unfortunately it is not evident who stores what, how and where. There is for instance not much information about whether they are creating a dedicated blockchain or use an existing one.
The Scienceroot project \cite{tintas_scienceroot} was launched in 2018 with the intention to create an on-chain economy around the publishing system using a reward token called Science Token (ST), which is deployed on the Waves blockchain. They also created or attempted to create an academic journal that ties into their economy. Pluto\cite{kang_pluto} is a blockchain-based platform for academic publishing that supports peer review, open access and micropayments. ARTiFACTS\cite{artifacts_team} is a project that aims to create a blockchain-based platform for scholarly research that enables researchers to create a permanent, time-stamped record of their various items that support their research such as data sets, images, figures etc. 
PubChain\cite{pubpub_team} is a project that aims to create a decentralized open-access publication platform that combines a funding platform with decentralized publishing. Like Scienceroot, it has its own token coincidentally also called Science Token (ST), which is used to exchange funds, store articles on IPFS and store their content identifiers on the blockchain. They also plan to integrate crowdfunding through their marketplace. TimedChain \cite{timedchain_team} is a project that aims to create a blockchain-based editorial management system that organizes manuscripts by publishers, authors, readers and other third parties. EUREKA \cite{schaufelbuhl2019eureka} is a project that aimed to create a blockchain-based peer-to-peer scientific data publishing platform with peer review, open access and micropayments. It was developed by the team behind ScienceMatters, an existing open access publisher that conducts triple-blind peer review. EUREKA also aimed to provide a blockchain-based rating and review system that allows readers to evaluate the impact of published articles. It is, however, not any more maintained. 
The Open Science company Desci Labs is developing a project called Desci Nodes \cite{descilabs}. Similar to Scienceroot, DeSci Nodes is a tool for creating research objects, which are a type of verifiable scientific publication that combines manuscripts, code, data, and more into a coherent unit of knowledge.  
The 2018 "nature index" article \cite{brock2018} entitled "Could Blockchain Unblock Science?" focusses on the question of how blockchain could be used to improve the process of current science publishing. Brock also mentions that data edits could be made permanently visible, which alludes to the idea of securing continuous editing in an immutable and consensual manner. He also developed and deployed the Frankl, which is an open source blockchain-based publishing platform \cite{brockOpenScience2018}.
Further insights into the landscape of blockchain-based solutions for scientific publishing are provided in \cite{mackey2019framework}. Apart from providing an overview of the landscape until 2019, they propose a governance framework for scientific publishing based on a consortium blockchain model. Some of those solutions aim to make the reward structure more open and introduce on-chain reward systems \cite{calcaterra2018chain}.

When developing solutions for academic publishing, blockchain technology seems appealing because it yields effectively immutable, globally agreed data in an open and transparent way without the need for a single source of trust. However, one must not fall into the fallacy of searching for nails for a hammer. At the heart of the blockchain paradigm lies the idea of a consensus about a global unique truth. This is a very useful technology for fiat (e.g. printed money or cryptocurrency), which exists through a global consensus. However, research output is not a fiat currency. It is subject to conflicting theories, opposing views and possibly irreconcilable results. All of those drive the continuous process that is science. One may build solutions on top of a blockchain to allow for potentially conflictual editing, but this is not what it was designed for. Instead we suggest to make Lakat a base layer that satisfies the requirements for a publication system by design. A comprehensive study on the development of decentralized consensus mechanisms in blockchain networks, such as the work by Wang et al. \cite{wang2019survey}, which provides an in-depth review of the state-of-the-art consensus protocols from both the perspective of distributed consensus system design and incentive mechanism design.
   
There are also some solutions that attempt to decentralize version control systems or anchor them in a blockchain. One of the most prominent examples of a decentralized version of a version control system with branches is git-ssb, a decentralized version control system based on the secure scuttlebutt protocol that allows for distributed version control without a central authority \cite{gitssb}. The Radicle protocol is another example, which is a peer-to-peer network for code collaboration that extends git with a networking protocol called Radicle Link \cite{radicle}. The project is governed through the RAD token, which is deployed on the Ethereum blockchain \cite{buterin2013ethereum}. Another project that explores ways to decentralize the storage of versioned data is Ceramic, a decentralized network for managing mutable information based on the idea of streams, which are append-only logs of JSON objects. The streams are anchored in a blockchain, which is used as a global ordering mechanism, and stored in a decentralized storage network \cite{ceramic}."

With the onset of large language models (LLMs) and AI agents that are capable of statistically extrapolating from a vast set of existing resources we are entering an era where some portions of the scientific research process can and should be outsourced to those models. AI agents should be able to take part in the process of scientific discovery. The impact and power of AI-aided or AI-generated research can be seen in multiple ways. For instance in the field of health and drug research, AI has helped improve the accuracy and efficiency of imaging \cite{Topol2019}, the interpretation of large datasets \cite{Libbrecht2015,Kraus2019,Holzinger2019} or the discovery of drugs\cite{AlQuraishi2019}. Moreover, the emergence of large language models has led to the development of autonomous scientific research capabilities, where these models can generate new hypotheses, design experiments to test these hypotheses, and interpret the results to draw conclusions, thereby playing a significant role in the scientific discovery process \cite{EmergentAutonomous2023}.

\subsection{Imre Lakatos}

The entire architecture of Lakat is heavily inspired by concepts developed by the Hungarian philosopher, Imre Lakatos. In an attempt to contribute to the demarcation problem \cite{popper1959logic,lakatos1978falsification,feyerabend1975against,pigliucci2013demarcation} that was prominent in the field of philosophy of science during Lakatos' times, he developed the concept of a \textit{research programme} \cite{lakatos1980methodology, worrall1980methodology, gavroglu1989imre}, also called \textit{Lakatosian research programme}, to avoid confusion with the colloquial use of the former term. The demarcation problem asks about the criteria that distinguish science from 'pseudo-science'. Lakatos develops his theory on the grounds of a process-oriented account of science. So rather than saying that this or that monolithic bulk of work or set of statements is or is not scientific, he posits that this distinction can only be made on the grounds of processes of theoretical amendments to an existing corpus of statements. He distinguishes between progressive and degenerative amendments depending on whether they strengthen the programme's predictive power. For Lakatos a research programme consists of a \textit{hard core}, which is a set of constituting assumptions, axioms as it were, that capture the essence of a research endeavor and a \textit{protective belt} of auxiliary hypotheses. The key ideas that the Lakat-architecture takes from the concept of the Lakatosian research programmes are threefold: 1) The pluralism of various research undertakings. 2) The process-orientation 3) The distinction between a core and a protective belt. At the heart of these foundational concepts lies the idea that science lives through arguments, differences and discourse. The input of Lakatosian concepts into Lakat can then be described as follows: A research programme corresponds to a branch or a set of branches to which researchers contribute changes or amendments. There is no single master branch, but rather every research programme has its own branch or set of branches. Conflicts with other branches or even within the same branch are an important aspect of Lakat and can be the source of progress (c.f. progressive amendments in Lakatosian research programmes). A programme can maintain a set of feature branches that support the core branch. These side branches behave like a protective belt.  

\subsection{Overview}

With Lakat, we propose a manifestly pluralistic, process-oriented, and conflict-oriented architecture for the continuous integration of publications, with a primary use case of research publications. In this way Lakat becomes a living document. At its core, the architecture consists of a linked data structure that resembles a DAG, where the main objects are branches. This data structure facilitates collaborative work in much the same way as git does. Branches may be thought of as the analogue of a journal in traditional publishing. The role of journal editors is covered largely by branch contributors. Branches are chains of blocks that contain submissions. The addition of another block happens via a proof of peer review, where the peers are the contributors to that branch. In that sense branches resemble blockchains with blocks consisting of submitted changes instead of transactions. As a consensus mechanism we discuss a solution that combines a proof-of-review at branch-level, a local (i.e. involving just branch-contributors) consensus rather than a global one, with a new finality gadget called Lignification. The review process is open. In a first version of Lakat the identities of the reviewers and the creators of the reviewed content are disclosed, however we wish to migrate to a weak\footnote{Weak is to be understood in the sense that both parties may choose to reveal their identity.} form of a double-blind protocol leveraging zero-knowledge proofs, where each party may reveal their identity.
Data is content-addressable and conforms to the IPLD CID format \cite{ipld2022}. Storage is handled by a networking component in Lakat, which delegates the bulk of data storage to a selection of other storage providers, including decentralized storage networks such as IPFS \cite{psaras2020interplanetary, guidi2021data, trautwein2022design}, storj, and others. This improves resilience and longevity. 

We emphasize that the main contributions of this paper are the high-level ideas for an architecture of a pluralistic process- and conflict-oriented peer-to-peer publishing system. We put forth one possible data structure and protocol. However, we also want to make it clear that these specifications are work in progress.  

In the following, we discuss the individual elements of the proposed system and highlight their interaction. We start with the data structure in Section \ref{sc:datastructure}, which is the core of the system. There we introduce the objects of a data bucket, a branch, a 'submit' as well as storage related aspects such as the data-trie and the database. We also discuss non-persisted parts of the data structure, namely the branch requests. In Section \ref{sc:participantsandnodes} we discuss the participants of the system and in particular the concept of a branch contributor. Then in Section \ref{sc:protocol} we discuss the protocol that handles the broadcasting, the consensus mechanism via a proof of review, a new finality gadget called Lignification and also how branches can be created, modified or operated on in this protocol.

%% file: datastructure.tex
\section{Data Structure}
\label{sc:datastructure}
\subsection{Bucket}
\label{ssc:bucket}
The most elementary data object is the \textit{bucket}. Each and every submitted item is submitted in a bucket: datasets, paragraphs, images and formulae are contained in buckets. These are examples of \textit{atomic buckets}, expressing the fact that they are the building blocks of the system. 
Instead of a folder structure, we solve the containment relation through designated buckets that we call \textit{molecular buckets} (like \textit{tree} nodes in git). The data part of those buckets contains merely an arrangement of atomic buckets. One may think of them as the analogue of an article, a book or some other curated content.

\begin{figure}[h!]
  \begin{center}
    \includegraphics[width=0.20\textwidth]{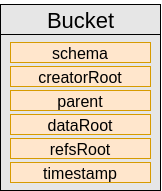}
\end{center}
 \caption{The most elementary type of data container is the bucket. It contains only immutable entries (orange), such as the \textit{schema}, the \textit{creatorRoot}, the \textit{parent}, the \textit{dataRoot}, the \textit{refsRoot} and the \textit{timestamp}.}
 \label{fig:bucket}
\end{figure}

Every bucket contains six entries: A \textit{schema}, a \textit{creatorRoot}, a \textit{parent}, a \textit{dataRoot}, a \textit{refsRoot} and a \textit{timestamp}. See Figure \ref{fig:bucket} for an illustration. Here and henceforth the word root refers to the root of a Merkle tree. We go through the entries in turn. The \textit{schema} contains details about the format of the data. For instance we have already mentioned that the data in the molecular buckets is formatted as an arrangement\footnote{The purposefully vague formulation of an 'arrangement' is due to the intention to keep that format flexible. One may think of this as an ordered list, but one might also consider further directives or clustering of content in a directed hypergraph.}. The \textit{creatorRoot} points to information about the creator of this bucket. In \textit{Lakat} contributors have one or many public-private keys and contributions are signed off with them (see Subsection \ref{ssc:accounts}). We wish to transition to a system where contributors only submit proofs of their contribution without revealing their identity (public keys etc). The \textit{parent} is the \textit{content identifier} of the parent bucket. For genesis buckets that would be 0. The \textit{dataRoot} is a content identifier of the data contained in the bucket. In future versions the schema could be absorbed into the dataRoot using the IPLD CID format. This would require a Lakat-specific codec. The \textit{refsRoot} points to all references made to other buckets within the data. This is necessary, since references to other buckets might be obscured inside the data-encoding. This is an analogue of a list of citations. The \textit{timestamp} records the time of inclusion of the bucket into the branch. It is important to note that we use Ethereum \cite{buterin2013ethereum} and some Layer2 block hashes as time stamps in our first version, since the local consensus is too weak to ensure that all participants are truthful to the time otherwise. Anticipating block hashes is close to impossible. One cannot change the data inside the bucket. One would have to create a new bucket that points to the original bucket via its parent entry.

\subsection{Branch}
\label{ssc:branch}

\begin{figure}[t!]
\begin{center}
\includegraphics[width=\textwidth]{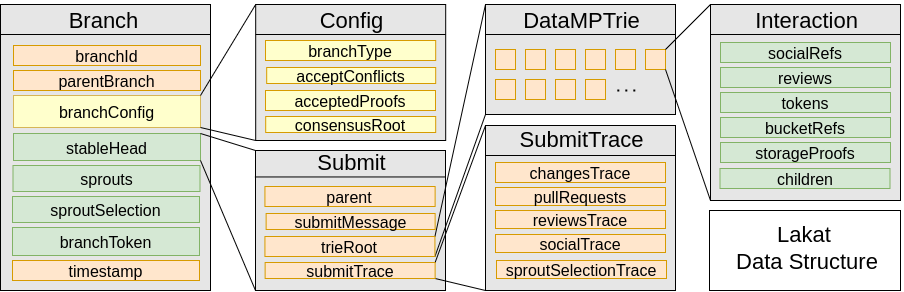}
\end{center}
 \caption{A schematic illustration of the branch object and its entries.}
 \label{fig:branchstructure}
\end{figure}

The central object type of Lakat is the \textit{branch}. See Figure \ref{fig:branchstructure} for an illustration. Branches represent journals or research communities. They share some properties with \textit{git}-branches and some with blockchains. Every branch contains an id, called \textit{branchID}, that uniquely identifies it. The immutable entries of a branch and the initial head are hashed to produce the branch identifier. The branch also points to a parent branch from which it originated. This entry may however be empty for a certain type of branch, namely the sprout (see below). The corresponding entry is called \textit{parentBranch}. This construction turns the set of branches into a linked data structure. In \textit{git} a branch is simply a pointer to the head commit. In blockchains one often encounters ids attached to the chain (so-called \textit{chainid}) to avoid issues when the consensus mechanism yields two different chains. At creation time the branch receives a \textit{timestamp}. The previous entries are all immutable. There are then four mutable entries, namely \textit{stableHead}, then the two consensus entries \textit{sprouts} and \textit{sproutSelection} and finally \textit{branchToken}. The stable head is a pointer to the latest stable submit. A \textit{submit} is a set of changes (see Subsection \ref{ssc:submit} on submits). One may think of it as the Lakat version of an article submission. It has similarities to a commit in git -- not only phonetically -- but also to a block in \textit{ethereum}. The addition of new submits works through a consensus mechanism called \textit{proof of review (PoR)} and \textit{lignification} (see Subsection \ref{ssc:lignification}). Also in this respect the branch behaves a lot like a blockchain. Every branch has it's own token, the \textit{branchToken}. It allows funding bodies to fund a particular branch. Token logic is not handled by \textit{Lakat}. Instead this entry essentially points to proofs of transactions on a blockchain where the respective token lives. The purpose of the integration of tokens is to create an incentive layer on top of Lakat, because (unfortunately) \textit{humans} as well as \textit{AI} do not work without incentives. The branch also carries configuration metadata, stored in \textit{branchConfig}. It points to information about the branch type, whether merge conflicts are accepted (see Subsection \ref{ssc:submit}), the consensus rules and the proofs that are accepted, such as proofs of token transfer or proofs of time. We use timestamps from latest blocks on various blockchains as proofs of time (see \cite{gipp2015decentralized} and also  \href{https://opentimestamps.org/}{opentimestamps.org}). The branchConfig's mutability is more constrained than that of the stable head (See Subsection \ref{ssc:configchange}).
Finally, we envision a way to extend the config schema. This would be done by an additional entry that points to a \textit{schema bucket}, where the schema for the config is defined. An empty entry would signify the use of the default schema.

There are three types of branches: \textit{proper branches}, \textit{sprouts} and \textit{twigs}. The branch type is stored in the branch config and can be changed under certain conditions. 
\textit{Proper branches} can only be modified through the local consensus mechanism (see Subsection \ref{ssc:localconsensus}). They point to a set of sprouts, which helps with the process of producing stable heads in the proper branch. Proper branches cannot be changed to any other branch type. A \textit{sprout} is a short-lived branch that is solely used to grow proper branches. Sprouts behave a bit like ommers in the ethereum protocol in the sense that they are contestants to produce the next stable head. They do not have an empty parent branch entry. Sprout branches point to an empty set of sprouts themselves. The sproutSelection contains all the sprouts that are rooted in it. The branchToken entry is empty. The stableHead is immutable. There is only one way to modify the sprout, namely indirectly when it turns into a proper branch during the lignification process (See Subsection \ref{ssc:lignification}). Once a sprout turns into a proper branch the parent branch entry is filled with the id of the branch that it is rooted in. Finally, a \textit{twig} can be thought of as a little feature branch. Twigs can be modified through submits by \textit{contributors} of the twig (See Subsection \ref{ssc:contributors} for more information on contributors) or through merges. However, the process of merging into a twig does not need to go through the consensus mechanism of proper branches (See Subsection \ref{ssc:localconsensus}).

In this paragraph we merely introduce some nomenclature. We distinguish between \textit{core} and \textit{belt} branches, which correspond to \textit{this} and \textit{other} in git. These are not intrinsic properties of branches, but denote the role they play during a merge. Lakat only has one type of merge. The core branch will be updated and the belt branch will not (see Subsection \ref{ssc:branchops} for information on merges). A branch may be a core with respect to one merge and a belt with respect to another merge. This terminology originates in the core-belt dichotomy of Lakatosian research programmes. There is a further distinction that is purely conceptual and is not manifested in the technical specification, but in the nomenclature. We distinguish a \textit{derived branch} from a \textit{seedling branch} in that the seedling branch has a \textit{singularity submit} without a parent (See Subsection \ref{ssc:submit} for information on submits). A singularity submit corresponds to the genesis block in a blockchain. We invoke here a cosmological metaphor rather than a biblical one. The seedling branch has no parent branch and the corresponding entry points to zero. A derived branch on the other hand has a parent branch that it points to. We say that the derived branch is \textit{rooted} in the parent branch. The \textit{root} of a derived branch is the last submit in the submit history that is also in the history of the parent branch.

\begin{figure}[b!]
  \begin{center}
    \includegraphics[width=1.0\textwidth]{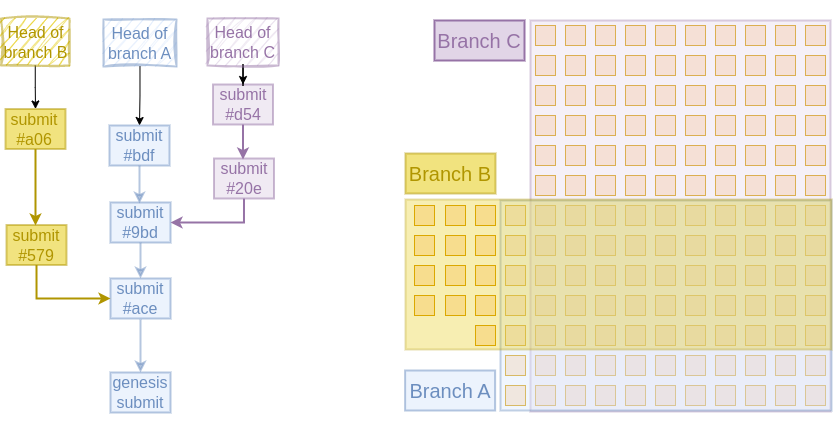}
\end{center}
 \caption{A schematic illustration of the two main objects: The branch and the data buckets. A branch typically references multiple buckets and any bucket may be referenced by many branches.}
 \label{fig:branchbucketrel}
\end{figure}

We also note that there are various levels at which Lakat can be viewed as a graph, going from high level to low level. At the level of the branches one can form a graph $\mathcal B$, where a branch is a node and a directed link from one branch $A$ to another branch $B$ means that $B$ is the parent of $A$ or that $B$ is merged into $A$ (See Subsection \ref{ssc:branchops} regarding merging). This directed graph is not necessarily a-cyclic, because $A$ can be rooted in $B$ and merge back into $B$, however if one excludes merges it is. At the level of the submits, a graph $\mathcal S$ can be created with the submits being nodes and a link can be drawn from a submit $q$ to $p$ when $p$ is the parent of $q$. This yields a directed acyclic graph (DAG). Finally at the level of the data buckets there exists a graph structure $\mathcal D$ induced by the parent reference inside the bucket. There is a graph homomorphism from $\mathcal S$ to $\mathcal B$, but not vice versa and there are no homomorphisms between $\mathcal S$ and $\mathcal D$ or $\mathcal B$ and $\mathcal D$. The lack of a homomorphism between the submit structure and the data structure indicates that these are two separate layers. The relation between the elementary bucket object and the higher level branch object is not simply a many-to-one relation. Different branches may share some data buckets. In practice one would expect that most of the data inside a branch is shared with at least one other branch. See Figure \ref{fig:branchbucketrel} for an illustration of this relation.

\subsection{Submit}
\label{ssc:submit}
Submits bundle up changes to the data with some additional metadata. Every submit points to a previous submit, the \textit{parent} submit. There exist \textit{singularity} submits that have no parent. The parent entry of those submits is zero. Like in Git \cite{chacon2014pro} or Ethereum \cite{buterin2013ethereum}, there is a field reserved for submit-specific data that we call \textit{submitMessage}. The change of the data within the submit is subsumed in \textit{trieRoot}, which is the root of the \textit{DataMPTrie}, a Merkle-Patricia-Trie that references the data state of Lakat (see Section \ref{ssc:datatrie}). The leaves of the trie are the data buckets. They have some resemblance with accounts in the ethereum state trie. Usually only a small part of the entire trie gets updated in a submit. Imagine the trie being all of wikipedia and a submit being just the creation of a new page or even just editing a page. Even though the bucket identifier is immutable it points to mutable entries. This is similar to ethereum, where the leaf nodes are immutable account addresses that point to mutable entries like amount of ETH, the contract storage data or the account nonce. The mutable entries in the case of Lakat are made up of information that is attached by other users to the bucket. It is information that is not intrinsic to the bucket. This includes \textit{socialRefs}, \textit{reviews}, \textit{tokens}, \textit{bucketRefs} and \textit{storageProofs}. 

The socialRefs entry resolves to tokens of appreciation, such as thumbs up or down -- the gold standard of social media user interaction. The reviews point to data buckets that contain a review or comments on the bucket in question. The tokens entry allows for the integration of tokens to data buckets. The bucketRefs are two collections of references to other buckets. The first collection is immutable and contains all those other buckets that are referenced inside the bucket data. This second collection is mutable and consists of all those molecular buckets that the atomic bucket is part of. This is a reverse registry that can be understood as how much a content has been reused. There is no analogue in classical publishing. StorageProofs are a ledger of timestamped proofs of storage for the bucket.  

There are some submits with a specific structure. These are the \textit{pull requests} (see Subsection \ref{ssc:por}) and the merge submits (see Subsection \ref{ssc:branchops}). The pull request  contains at least one context bucket, called the \textit{review container}, that references all the subsequent reviews. It also leaves a trace of the pull request in the submitTrace. The merge submit contains 
all the data buckets of the belt branch and it points to the merged branch id in the submitTrace. 

In Lakat conflicts are at the heart of the protocol. They are cherished as the source of progress and sets Lakat apart from conventional publishing systems. We provide a clear definition of a conflict. A \textit{submit conflict} with respect to a branch $\mathfrak B$ is a set 
of three submits $\pi$, $s_1$ and $s_2$ where $\pi$ is the parent of both $s_1$ and $s_2$ and all three are included in $\mathfrak B$. We denote this 4-tuple by $(\mathfrak B, \pi, s_1, s_2)$.
A submit that creates a submit conflict is called a \textit{conflicted submit} and a submit that does not create a submit conflict is called \textit{conflictless submit}.
A \textit{merge conflict} is a submit conflict that arises from a merge submit. Depending on the branch configuration (see Subsection \ref{ssc:branch}) merge submits may or may not bring about merge conflicts.

\subsection{Data--Trie}
\label{ssc:datatrie}

The data buckets as well as the mutable information attached to them can be looked up with the help of a Merkle-Patricia trie, called the \textit{DataMPTrie}. This is cryptographically secure and very useful when resolving the information attached to buckets inside of an article. The keys that are stored in the trie are truncated versions of the content identifiers of the data buckets. And the values are the mutable entries attached to the buckets. To look up the bucket data itself one uses simply the content identifier of the bucket. Storage is handled separately (see Storage in Section \ref{ssc:storage}).
We propose to use a modified Merkle-Patricia trie -- very similar to the one used in Ethereum -- with four types of nodes: null nodes, leaf nodes, extension nodes and branch nodes.
The data at each node is serialized and hashed. The specifics of this encoding are yet to be specified. One may use any of the existing IPLD-formats. The encoding should have the property that data lists with a lot of empty entries are serialized in a very compact way to save space. Many data items in Lakat have a lot of empty fields. A bucket without any interaction information is mostly empty fields. Twigs and sprouts have many empty fields as well. The leaf-nodes (in the trie) are special in this respect, because the hashing uses a salt that equals the content identifier of the bucket. Why do we need a salt at all? When a data bucket is published it doesn't have any information attached to it, so without the salt all new data buckets would have the same hash, which is not desirable.

\subsection{Storage}
\label{ssc:storage}

The data is stored in key-value databases and is content addressed in the sense that the key equals the CID of the data. We would like to use the IPLD standard for linked data \cite{ipld2022}. A piece of data can be stored with multiple protocols. A contributor may also choose to store the information on their own machine of course. The more branches point to a piece of data and the more subsequent submits rely on it the more important the persistence of that data becomes.   
The idea is that the availability, the longevity and the redundancy of data will scale with its importance in a self-organized fashion. A branch with many contributors will make sure to have the storage well secured and also well distributed. A newly created branch on the other hand needs to broadcast its creation (see Subsection \ref{ssc:branchrequests} for branch creation broadcasting) to allow for distributed storage  and attract contributors to ensure decentralized persistence of its data. 
This has two advantages. 1) Data that is pointed at by many branches is highly available and more redundant. 2) One cannot attack the system by creating lots and lots of branches. To prove that a certain data bucket has been stored, i.e. pinned, that proof is attached to the mutable information of the bucket in the \textit{storageProofs} entry. There are a few more constraints about storage and pinning. It should be encouraged that every data bucket belongs to at least one molecular bucket so that there are no buckets without a context. Thus when a new data bucket is submitted the submission won't be accepted unless it is present in at least one context bucket.

When a new branch is created the data is initially just stored by the branch creator, but broadcasted through the network. Some nodes may pick up the data and store it as well. The branch creator may also choose to pin the data bucket in a certain storage system. Data that is close in the branch data structure is also close in the storage system. This is a very important feature of Lakat. It allows for a very efficient retrieval of data. The storage of data pertaining to a branch can be rewarded in branch tokens. This is not a feature of Lakat, but may be added on top of it to incentivize storage. There may also be a market for storage, where branch creators can buy storage space for their branch data. This is also not a feature of Lakat, but may be added on top of it.

\subsection{Branch--Requests}
\label{ssc:branchrequests}

\begin{figure}[h!]
  \begin{center}
    \includegraphics[width=0.25\textwidth]{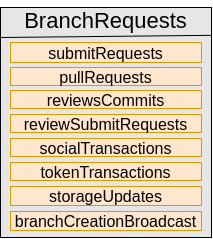}
\end{center}
 \caption{The eight channels of the branch requests. Branch requests are the staging area of branches. Each channel has a limited capacity.}
 \label{fig:mempool}
\end{figure}

Every branch has its own staging area, where any type of branch interactions is waiting to be included. This is called the \textit{Branch--Requests}. It is similar to the mempool in ethereum. Everyone participating in the branch (See Subsection \ref{ssc:contributors}) may receive branch interactions from users and broadcast them to the network. Here we refer to a \textit{client} as a piece of software that is yet to be written, which interacts with the network. A \textit{light client} is a client that is not capable of doing branch operations, but is capable of receiving and broadcasting branch requests. Inclusion of requests into the branch, however, requires more (see Subsection \ref{ssc:por}). There are eight channels in the branch requests (see Figure \ref{fig:mempool}): \textit{submit requests}, \textit{pull requests}, \textit{review commits}, \textit{review submit requests}, \textit{social transactions}, \textit{token transactions}, \textit{storage updates} and \textit{branch creation broadcast}. The requests inside the Branch requests are not permanently stored as part of the protocol. Requests are kept for as long as any of the branch contributors keeps track of them. That is where the similarity to the mempool stems from.

Every channel in the Branch Requests has a certain capacity. In particular this aims to prevent that one channel clutters the entire pool of requests, which might happen if the capacity was channel-independent. All requests or broadcasts are serialized. 
Submit requests contain serialized versions of the data buckets that are requested to be added to the branch. Pull requests are simply notifications from other branches that seek reviewers. Only by means of a pull request can contributors from the target branch be allowed to make modifications to the requesting branch (see Proof of Review \ref{ssc:por}).


%% file: contributions.tex
\section{Participants}
\label{sc:participantsandnodes}

\subsection{User Identity}
\label{ssc:accounts}

We propose to use an identity management system that ties in with some of the existing decentralized identity schemes and allows for the integration of multiple signing methods. For the first version of Lakat we propose to use 3id as the identity management system. 3id is a W3C compliant decentralized identity management system that allows for the integration of multiple signing keys. The identifier for 3id within the did system is did:3. It relies on a mutable document type in the ceramic network, called a stream. In the future we would also like to reduce identity to the ability to prove the submission of content without exposing further information about the identity using zero knowledge technology without a trusted setup. This would allow for a double blind review system. In order to publish content or send messages into the network a user needs to have an identity, which at this point is a did:3 identity registered in the ceramic network. The private keys in the did are used to sign messages and to prove authorship of content. The keys are also used to sign messages that are used to propose new states of a branch.

\subsection{Contributor}
\label{ssc:contributors}
Every branch has \textit{contributors}, or rather contributors have branches. A contributor is an account that can prove to have contributed to a given branch. There are four types of contributors for any given branch: \textit{content contributors}, \textit{review contributors}, \textit{token contributors} and \textit{storage contributors}. A content contributor can prove to have submitted to the branch. A review contributor is someone who can prove to have pushed reviews to the branch (see Subsection \ref{ssc:por} for information on proof of review). A token contributor is someone who can prove to have deposited funds into the branch. A storage contributor is someone who can prove to store data of that branch. Being a contributor means that you have to prove your contribution for the submits from the root submit of the branch till the current stable head.
How does the set of contributors change during a merge? What is the relation between the contributors of two branches before the merge and after? When the belt branch is merged into the core following a pull request, then the new set of contributors is simply the union of the two branches (see Subsection \ref{ssc:branch} for the terms core and belt). That holds for all contributor types. When there is no pull request preceding the merge the contributors of this branch are unaltered.
The main idea behind the concept of contributors is derived from the mutability, governance and autonomy of branches. Branches can only be modified by their contributors. This attempts to preempt attacks on branches.

\subsection{Contribution}
\label{ssc:contribution}

A contribution is any type of interaction with the branch. There are four types of contributions: \textit{submit}, \textit{review}, \textit{token} and \textit{storage}. A submit is a contribution that adds data to the branch. A review submit is a contribution that adds a review to a branch. A token submit is a contribution that adds a proof of a token creation, token mint or token transfer to the branch. A storage contribution is a proof about the storage of data contained in the branch, i.e. pointed at by submits of the branch. A contribution is always associated with a branch and a contributor. In the first minimal viable version of Lakat, we are planning to use zk-STARKs as a proof system. This is due to the fact that zk-STARKs do not require a trusted setup. We use Cairo programmes to generate proofs and point to their verification. The proof of contribution is a hash of the zero-knowledge proof of the contribution and its verification.

When a branch request is sent into the network it is being routed using the Kademlia protocol \cite{maymounkov2002kademlia} to the contributors of the corresponding branch. The backlog of requests is being shared and continuously broadcasted and updated by storage and content contributors using the libp2p library \cite{libp2p}. If the request has a payload that ought to modify the branch state, the receiving node checks the proof of contribution. If the proof is valid the node adds the contribution to the branch. If the proof is invalid the branch rejects the contribution. If the contribution is a submit and the submit is not valid it cannot be included in the branch.

%% file: protocol.tex
\section{Protocol}
\label{sc:protocol}

In this section we describe the Lakat protocol. We start with a high-level overview of the protocol and then go into the details of the individual components. Lakat is a shared key-value database of branches and data buckets together with a peer-to-peer protocol that governs the modification of this database. The modifications happen through submits to branches. The protocol needs to cover five functions: 1) Define a mechanism to construct new contributions \footnote{This is called block proposal for blockchains. Examples include proof-of-work or proof-of-stake}. 2) Broadcast information about requests and new branch modifications through the network. 3) Check the validity of the branch modification. 4) Define a strategy to finalize the state of a branch. 5) Incentivize contributors to propose modifications. 

Lakat proposes a local consensus mechanism that relies on the notion of branch contributors. In principle Lakat could be used with various consensus mechanisms at branch level, such as proof-of-work or proof-of-stake. However, we propose a new one that we consider more suitable for academic publishing. This mechanism combines three concepts: 1) The distinction between \textit{feature and production} branches 2) A \textit{proof-of-review} mechanism that is used to propose new states of the production branch 3) A finality mechanism that is used to finalize the head of a branch, which we call \textit{lignification}. The incentive mechanism is not built into Lakat, but may be added on top of it through the token handling at the level of the branch. Even in the current publishing business the incentives are outsourced to reputation, job promises and in some cases mere scientific curiosity. If anything, there is an anti-incentive to publish. In the following we describe the individual components of the protocol in detail.

\subsection{Networking}
\label{ssc:networking}

One of Lakat's components is an asynchronous networking protocol, where peers can enter and leave at any time. The state of the individual processes of each peer is communicated and updated through a gossip protocol. The gossip protocol is used to broadcast requests and branch modifications to the network. We use the Kademlia DHT for this purpose. In Lakat the gossiping network is used to store the information state of the individual peers. This includes the branch requests (see Subsection \ref{ssc:branchrequests}) and the high-level information about the states of the branches that this peer keeps track of. This high level information consists of the branchId, the parentBranch, the branchConfig (branchType, acceptConflicts, acceptedProofs, consensusRoot), the stableHead (parent submit, submitMessage, trieRoot, submitTrace), the sprouts, sproutSelection, branchToken and timestamp (see Subsection \ref{ssc:branch}). What about the bulk of the data, namely the data trie with all the data buckets and their respective interaction information, and the trace of the stableHead? That is optionally outsourced to other protocols. The protocols are then part of the multihash. They could include e.g. IPFS \cite{psaras2020interplanetary, guidi2021data}, filecoin, Urbit\footnote{We also consider building on top of the Urbit OS using linedb\cite{linedb} as a key-value storage and networking solution} or if that peer chooses to do so, it could also store the data locally in the hash table. In Kademlia proximity of data is measured in proximity of its content identifier. In future releases of Lakat we propose to tweak the proximity such that data stored on the same branch is also close to each other in the routing distance. We are planning to use the libp2p library as a basis of the networking protocol \cite{libp2p}. It is a modular networking stack that uses Kademlia.

\subsection{Local Consensus}
\label{ssc:localconsensus}

Who decides which content will be added to a branch? In Lakat there is no global notion of what counts as science and what does not. There is only a local notion, the details of which is the subject of this section. A global consensus mechanism seems to be a good fit for a ledger that keeps track of values that are or ought to be globally agreed upon, i.e. for values that exist qua their global agreement. In contrast to money transactions, the global scope seems ill-fitted for the publication of research content. In our view this requires a local form of consensus. In the context of Lakat, the scope of the locality is at the branch-level. 

What does a branch-level scope mean? This means that the scope is constrained to the \textit{contributors} of a branch. Every branch has a history of submits and is \textit{rooted} in some parent branch or is itself a \textit{seedling} (See Subsection \ref{ssc:branch} regarding roots and seedlings). In either case there is a set of contributors to every branch between its root and the current head. Any actor, human or AI, who can prove to have contributed content in any of the branch's submits counts as a contributor (See Subsection \ref{ssc:contributors}). 

Branch contributors form the basis of the consensus mechanism. We entrench this deep into the protocol by allowing only branch contributors to make changes to the branch that they are contributing to. This design choice also keeps potential attackers from pushing unwanted content to a branch. Lakat does not make statements about what counts as science and what not. What counts as a legitimate scientific contribution purely emerges through the local consensus. One contribution that is viewed as being unfounded or unscientific for one branch might be viewed the opposite on another branch. In some sense this reflects a Feyerabendian approach \cite{feyerabend1975against}. It gives space for pluralism and allows for the organic selection of branches with possibly differing criteria on what counts as valid output. There is, however, a convergence in accepted method and output expected to emerge within a branch and also in branches that are close to each other. Branches that are close have branched off recently and possibly disagree more on technical grounds than on methodic grounds or they are simply feature branches that are to be merged back into the main branch soon. There is an overall incentive to merge branches, derived from the persistence of the data and the value of the token. 

The local consensus paradigm is governing amendments to branches, both to twigs and to proper branches. Sprouts on the other hand are just auxiliary objects that cannot be modified directly and are thus not amenable to a consensus mechanism. The consensus paradigm for twigs simply states that any branch contributor can push submits to the twig whereas merging into a twig requires a certain fraction of contributors to agree (See Subsection \ref{ssc:branch} for twigs and Subsection \ref{ssc:twiglocalconsensus} for consensus on twigs). For proper branches the local consensus takes on a different form. It is divided into proof of review (See Subsection \ref{ssc:por}), broadcasting and lignification (See Subsection \ref{ssc:lignification}).

\subsection{Feature Branches}
\label{ssc:twiglocalconsensus}

Twigs are meant to be used for rather quick iteration. They behave like feature branches. Here is an example where twigs are expected to be used: If a contributor, human or AI, would like to add content to a target branch, say an article or some modifications or both, it creates a feature branch rooted in the target branch which subsequently goes through the proof-of-review consensus mechanism (see Subsection \ref{ssc:por}). Typically the number of content contributors on a twig will be low. Maybe a single author or a small group of authors, as it is the case for article publications. In order to not compromise the momentum and the quick iteration both content contributors and review contributors (if there are any) can push to the branch directly. Merges can also be pushed, but require a fraction of approvals of the content contributors. The fraction is determined in the config of the twig. 

\subsection{Proof of Review (PoR)}
\label{ssc:por}

Before a branch can be merged into a proper branch it needs to undergo a review. Table \ref{tab:reviewProcess} summarizes the steps.
To start the review process an \textit{issuing branch} creates a pull request from a \textit{requesting branch} to a \textit{target branch}. The pull request is a submit with two properties: First it contains a newly created context data bucket, called the \textit{review container}, that will hold all the forthcoming information of the review. The submit may of course contain other buckets besides that. Second, it leaves a trace of the information about the pull request in the pullRequests entry of the \textit{submitTrace}, namely pointers to the review container, to the target branch and the requesting branch. In most cases the review happens on a twig, which acts as a feature branch. There the issuing branch and the requesting branch are identical, because the twig requests for itself to be pulled into the target branch. However, the requesting branch may also act as a proxy requester. This is the case when a proper branch rather than a twig seeks to be merged into a target branch. Since this intention itself must pass through the consensus rule of that proper branch, one would have to create a twig and include therein the proxy pull request. Once that twig is successfully merged into the actual requesting branch by passing the consensus, the review process can begin on that proper branch for it to be mergeable into the target branch. We call a pull request \textit{mature} once it is included in the requesting branch. In the most common scenario where the issuing and requesting branch coincide, maturity is immediate. 
Once a pull request becomes mature a message will be sent to the target branch  where its contributors are invited to review the requesting branch. The message is simply a reference to the pull request sent to the pullRequests channel of the target's \textit{branchRequests} (see Subsection \ref{ssc:branchrequests} for branch requests).
Any content contributor of the target branch who is not also contributing to the requesting branch can then become a \textit{review contributor} of the requesting branch. They must first publish a review commitment on the requesting branch. This makes them official contributors to that branch. It also helps to gauge general reviewers engagement prior to the actual review. This is helpful both for those who seek to merge and those who seek to review. It also increases accountability of the committing reviewer. Failing to supply a review after a commitment could be penalized via the social engagement.  
Committers publish their commitment in the reviewsTrace of the submitTrace. They cannot submit reviews without a prior commitment. Also, the identity of the reviewer is not public in the sense that the commitment solely contains a zero-knowledge proof that the reviewer is a contributor to the target branch (see Subsection \ref{ssc:contributors}).
Of course the reviewer may decide to reveal their identity and this may or may not be in line with the configuration of the target branch.

Reviewers then push review submits to the requesting branch. The submits just contain a proof of contributorship in the target branch. A review submit consists of the following: A bucket with a review, called a \textit{review item}. This bucket should reference all the data buckets that it has reviewed. In the respective interaction data (see Subsection \ref{ssc:branch}) of all those reviewed buckets a reference to the review item is stored within the reviews entry. Finally the review item gets referenced in the review container of the pull request. Updating the review bucket, as with any bucket update, consists of creating a new review bucket that points to the old one through the parent entry\footnote{In future versions of Lakat we wish to move to updates via deltas.}. The branchConfig of the target branch specifies the prerequisites for a merge. This consists of the minimum number of reviewers, a rule for acceptance and a minimum number of review rounds, which could be one by default. The rule of acceptance could be preset as well. For instance one could reject requests when a certain fraction of reviewers reject and accept when there are no rejections and specify some rule for the middle ground. Once all the requirements of the target branch are satisfied the branch is ready to be merged. 

How about merging branches that do not seek to be merged? This can be the case when trying to merge the newest developments from a remote branch. This case is in fact already covered by the respective consensus mechanisms of twigs and proper branches. Merging into twigs requires a fraction of content contributors to agree (see Subsection \ref{ssc:twiglocalconsensus}). Merging into proper branches requires a pull request and subsequent reviews, so it is not possible to just merge other un-reviewed branches in the same way that one merges reviewed twigs or reviewed proper branches. Therefore, one would have to create a twig that merges the remote branch as a feature. It then requests to be merged and the merge undergoes a review.

\begin{table}
  \begin{tabular}{|p{0.025\textwidth}|p{0.3\linewidth}|p{0.595\textwidth}|}
  \hline
  \textbf{\#} & \textbf{Step} & \textbf{Description} \\
  \hline\hline
  1 & Create pull request & The issuing branch creates a request for the requesting branch (in most cases identical) to be merged into the target branch. A review container is created. \\
  2 & Maturity of the pull request & The pull request is included in the requesting branch (in most cases immediate)\\
  3 & Commitment & A content contributor of the target branch publishes a review commitment to the requesting branch. That makes them review contributors of the requesting branch.\\
  4 & Review & The review contributors create review submits that are referenced in the review bucket.\\
  5 & Completion & The number of review cycles and the coverage of the review meets the criteria of the target's branchConfig. The branch may be merged into the target.\\
 \hline
  \end{tabular}

  \caption{Overview of the Proof--of--Review (PoR) process}
 \label{tab:reviewProcess}
\end{table}

\subsection{Broadcasting and Lignification}
\label{ssc:lignification}

How are the reviewed pull requests bundled up and sequenced into a single proper branch? Why is the process important? How is the required attention bandwidth for this process kept to a minimum? In order to explain the Lakat answer to this question we first contrast it to the case of blockchains: Their transactions are bundled into blocks. They are then broadcasted across the network of nodes. When different blocks with the same parent are broadcasted, there will be conflicting versions of the blockchain state, which for a single source of truth is undesirable. In ethereum prior to the transition from the proof-of-work to the proof-of-stake these alternative versions were called ommers and were mostly the result of latency in the broadcasting, but of course also attacks or client-software issues. To make sure that a transaction has irrevocably been added into the blockchain one would have to wait for a few block confirmations. 

In Lakat we solve the issue through a process that we call \textit{lignification}. The idea is that amongst the potentially plentiful and conflicting versions of the new branch state eventually a new head will be chosen. This head is then called the stable head. The versions are stored as short-lived branches, called sprouts. The \textit{sprouts} entry of the branch points to them. Why is the process of choosing a successor to the stable head important? Here is an explanation: The branch is an object that is kept alive by an ecosystem of contributors. It could get hijacked by a group of bad actors who became branch contributors through a mal-reviewed pull-request. In principle, if this happened, the contributors that disagreed with this malicious onboarding could bail out by creating a new branch. However, this new branch would have to grow the reputation of its contributors anew, seek new storage providers, have a new branch token and would generally have to start from scratch. It might not even be an attack, but a disagreement in the community that leads to a branching.
Even though the process of finding a new stable head constitutes an important security measure for the branch, it should not create an overload of attention demanded from the target branch contributors. In most cases there will be no action required. But it is precisely those rare cases, where such a security measure becomes valuable. So one of the requirements for this process is that the branch production continues unambiguously when there is no interference from the community of contributors. In the following we introduce the process of broadcasting and lignification in more detail.

\subsubsection*{Broadcasting}

Henceforth we refer to our proper branch as \textit{core}. It functions as a production branch. Every proposed new merge submit could either become the stable head of core or become the first submit of a new (disagreeing) branch that is rooted in core. We refer to any of those new branches collectively as the \textit{belt} branch. Note that the \textit{core} branch may also become \textit{belt} for another branch. Merge submits carry in themselves the possibility of becoming the head of a new branch. Therefore we decided to "wrap" them into short-lived proto-branches, namely sprouts, whose respective heads are the merge submits. The process of broadcasting is as follows. A content contributor of core, let's call her Alice, creates a merge submit, which is a special kind of submit (see Subsection \ref{ssc:submit}). This submit is then wrapped into a sprout, which means that the head of the sprout is set to be the merge submit and the content contributors are set to be the union of Alice and all the contributors of the pull-requesting branch. Let's call this sprout \textit{S}. The branch information of the sprout becomes relevant if it eventually turns into a proper branch, a process which is discussed in the lignification part of this Subsection. The parent of the merge submit is the head of a branch \textit{B}, that is either the core or any of the sprouts upstream of the core\footnote{The sprouts entry of a proper branch keeps track of all the upstream sprouts, but depending on the last branch update may also contain outdated sprouts. In order to retrieve all upstream sprouts one may "walk" upstream using the sproutSelection entry, which only contains the immediate offspring sprouts of a given branch.}. Alice chooses \textit{B}, so she decides where to root the new sprout. If she decides to point to a branch that is already pointed at, there will be a conflict. The new sprout \textit{S} -- or rather its branchId -- is then added to the sproutSelection entries of \textit{B} and the sprouts entry of the core (which might coincide with \textit{B}). The new state of core is then broadcasted to all contributors of core. Note that the new state of core might have received more updates than just the modification of the sprouts or sproutsSelection entries. There can also be further modifications resulting from the lignification process (see next part). The changes, i.e. creation, of the sprout branch \textit{S} are also broadcasted to its contributors. 

\subsubsection*{Lignification}

Once a given submit is the new stable head of core or of belt, it cannot be revoked. We say it is \textit{lignified}. The conversion of a previously flexible object into a rigid amendment of a branch has similarities to the process of lignification in botany. 
The decision about the stable head is not made immediately, but there is a period of time where it can still be revoked and deferred. This time is called \textit{lignification time}. As mentioned above, the objects that we make decisions about are not the merge submits themselves, but the sprouts that contain them. If there is only one sprout available after the lignification time, then the decision is clear, namely that the submit contained in that sprout becomes the new stable head of core and no action is required. However, there may be multiple sprout options. In this case, we propose to have a deterministic rule that singles out one sprout and we suggest the possibility of vetoing the default deterministic choice. This minimizes the need to vote each time multiple options arise, but more importantly it reduces the attack vector for people to bring branch growth to a halt by proposing alternate -- yet still reviewed -- merge submits. Vetoing is possible throughout the lignification time. Any branch contributor may register a veto to any of the vying sprouts and therefore against the default sprout. In case that a veto is registered the sprout in question has a chance to provide the next stable head. 
Once a veto is registered, the content contributors can bring in their votes on the rivaling sprouts. After a period of time, called the \textit{engagement time}, the winning sprout will provide the new stable head and the other sprouts can turn into peripheral proper branches rooted in core. Like with blockchains, the state of Lakat does not change by itself, but only through transactions (See Branch Requests \ref{ssc:branchrequests} and Submits \ref{ssc:submit}). This means that only when a new submit is broadcasted can the state of a branch be updated. Furthermore, a branch may only be updated if it is the target of a transaction. If the transaction is targeting core, then peripheral branches cannot be updated and vice versa. As a consequence those ousted rival sprouts do not turn into their own branches immediately, but only once a transaction targets them. Some of them may never turn into proper branches at all. Apart from the lignification time and the optional engagement time there is a time allowing for latency issues in broadcasting, called the \textit{broadcastingBuffer}. This ensures that the timestamped vetoes or votes are broadcasted and thus recorded before the stable head is irrevocably fixed.

\begin{figure}[h!]
  \begin{center}
    \includegraphics[width=0.75\textwidth]{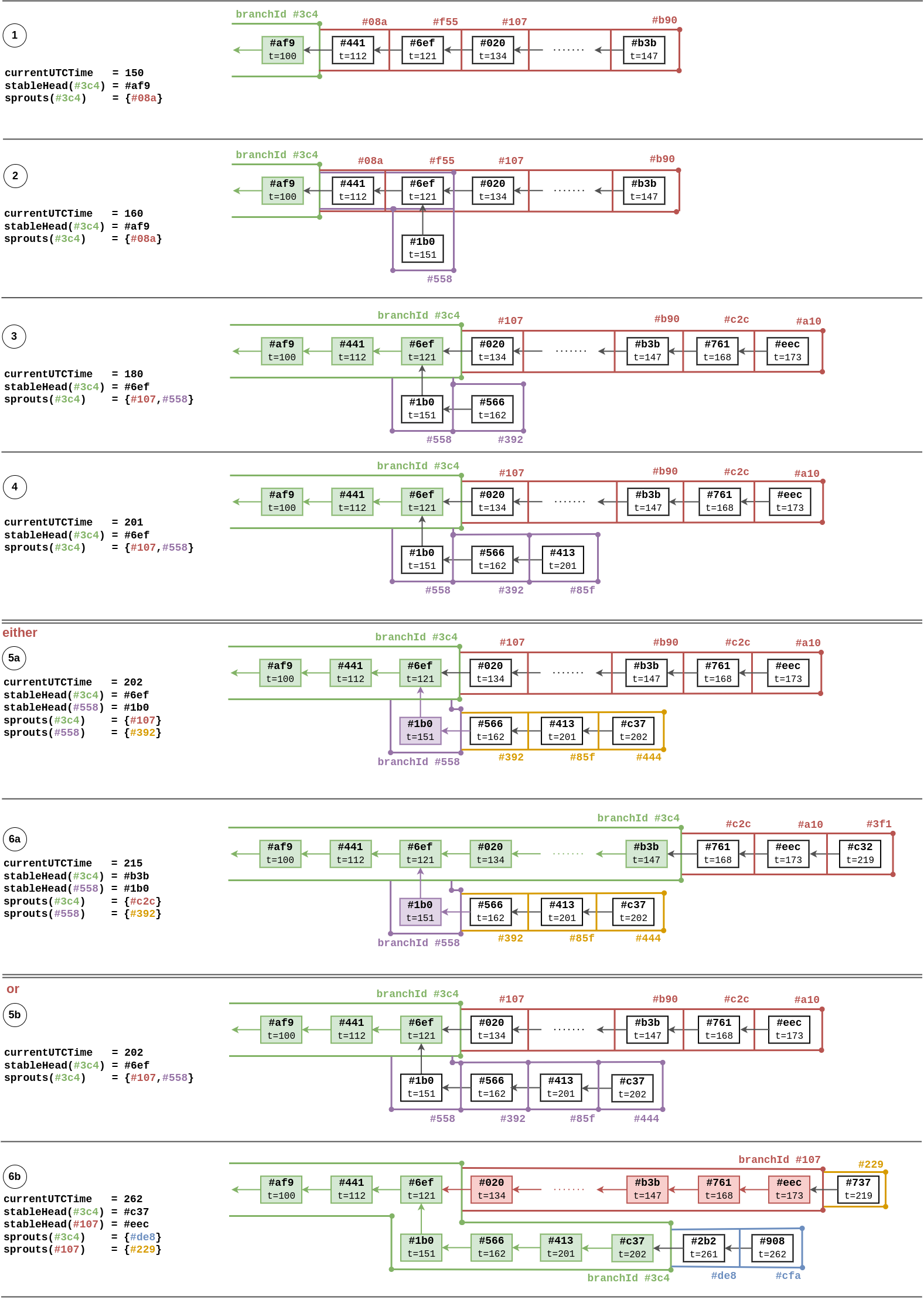}
\end{center}
 \caption{Example of a branch lignification with \textit{broadcastingBuffer} $=1$, \textit{lignificationTime} $=50$ and \textit{engagementTime} $=60$. Whenever a new mergeSubmit is added the lignification algorithm \ref{alg:lignification} runs and updates the stable head. The updates 1 to 4 are unambiguous. But then the target branch has two competing sprouts. The default sprout is \textit{\#107} and the other one is \textit{\#558}. Without any veto \textit{\#107} will deliver the next stable head of branch \textit{\#3c4}. This is the scenario 5a. The veto time plus \textit{broadcastingBuffer} have passed and a new mergeSubmit \textit{\#c37} inside sprout \textit{\#444} triggers the lignification algorithm so that the loosing branch \textit{\#558} becomes lignified and rooted in \textit{\#3c4}. In 6a the mergeSubmit lignifies the target branch \textit{\#3c4}. Its head has advanced to \textit{\#b3b}. In scenario 5b a veto had been registered for \textit{\#558} in the sproutSelection entry of the target. If the voting turns out to be in favor of \textit{\#558}, the lignification process will grow the target branch in that direction (c.f. Step 6b).}
 \label{fig:bufferbranches}
\end{figure}

Due to the time between successive transactions it is quite possible that the state of the core, in particular its stable head, needs to be updated. Maybe the veto time or the voting time between vying sprouts has passed or maybe there are no rivaling sprouts and the stable head simply needs to be advanced. The pseudo-code in the Lignification Algorithm \ref{alg:lignification} outlines the iterative procedure that advances the stable head on each new transaction. It is worth noting that also the sprouts entry and the sproutSelection entry of core get updated by pruning and replacement respectively.
An illustration of the lignification process is also shown in Figure \ref{fig:bufferbranches}.

In practice the broadcasting and lignification can be automated by a script so that it requires less cognitive bandwidth. The script would choose a content contributor of core at random and broadcast collect all the pull requests that meet the merge-requirements from core, then create one or more merge submits from them, go through the lignification process and broadcast the result. Only in the case when there are disagreements would a manual interference be required.

\begin{algorithm}[h!]
\caption{Lignification -- Advancing the stable head of the branch}
\begin{algorithmic} 
\REQUIRE coreBranch, mergeSubmit, broadcastingBuffer, lignificationTime, engagementTime 
\STATE downstreamBranches $\leftarrow$ branches downstream of coreBranch: [coreBranch, ..., sproutOf(mergeSubmit)]
\STATE referenceBranch $\leftarrow$ coreBranch \comment{/* referenceBranch may later be core or belt branch */}
\FOR {i in 1 ... (downstreamBranches.length - 1)  \comment{/* indexing starts at 1 */}}
  \STATE currentBranch $\leftarrow$ downstreamBranches[i]
  \STATE childSprout  $\leftarrow$ downstreamBranches[i + 1]  \comment{/* always exists */}
  \IF {all currentBranch.sprouts are within lignificationTime time (plus broadcastingBuffer)}
    \RETURN
  \ELSE
    \IF {There is a veto against defaultSuccessor(currentBranch) and voting has finished}
      \STATE \comment{/* engagementTime is over (plus lignificationTime plus broadcastingBuffer) */}
      \IF {childSprout does not win the vote}
        \STATE \comment{/* doesn't participate (not defaultSuccessor or not part of a veto) or participates and doesn't win */}
        \STATE childSprout becomes a peripheral branch rooted in referenceBranch.
        \STATE referenceBranch $\leftarrow$ childSprout.
      \ELSE 
        \STATE \comment{/* childSprout wins the vote */}
        \STATE set the stableHead, sproutSelection and sprouts of referenceBranch to those of childSprout
      \ENDIF
      \STATE \comment{/* childSprout may or may not be defaultSuccessor. Both cases are covered. */}
    \ELSIF {There is a veto against defaultSuccessor(currentBranch), but voting has not finished}
      \RETURN
    \ELSE 
      \STATE \comment{/* There is no veto against defaultSuccessor(currentBranch) */}
      \IF {childSprout is defaultSuccessor(currentBranch)}
        \STATE set the stableHead, sproutSelection and sprouts of referenceBranch to those of childSprout
      \ELSE
        \STATE childSprout becomes a side branch rooted in referenceBranch.
        \STATE referenceBranch $\leftarrow$ childSprout.
      \ENDIF
    \ENDIF
    \STATE \comment{/* Note that the referenceBranch may have changed. */}
  \ENDIF
\ENDFOR
\RETURN
\end{algorithmic}
\label{alg:lignification}
\end{algorithm}

\subsection{Branch Config Changes}
\label{ssc:configchange}

The branch config contains configurable metadata such as the branch type, a flag that can be set to allow only conflictless submits (see Subsection \ref{ssc:submit}), then the accepted proofs (e.g. proofs of storage, proofs of contributorship,  proofs of token transfer) and also the parameters that determine the consensus (e.g. the number of reviewers needed in the proof of review algorithm). These entries have constrained mutability. They require a merge rather than a plain commit to take effect. For a twig this means that the consensus mechanism for a twig needs to be met, i.e. a config-specific fraction of contributors need to approve the merge. For a proper branch this means that the config change needs to go through the proof-of-review (PoR) consensus mechanism (see Subsection \ref{ssc:por}). We envision that in some future release there will be a default schema for the config, but that this schema may be altered through schema buckets to which the schema is pointing.


\subsection{Branch Operations}
\label{ssc:branchops}

\subsubsection*{Creation} 
The first branch operation is the creation. There are \textit{genesis creations} and \textit{rooted creations}. As the name suggests, the genesis creation is a branch that does not have ancestral submits. This is similar to blockchains or git, which have a block or submit without a parent. However, unlike those, Lakat allows for multiple genesis creations. Anyone can at any time create a new genesis branch, which is either a twig or a proper branch. A genesis creation requires the creator to set the branch config. Optionally the creator may also specify a branch token. 
On the other hand a rooted creation is a branch creation in which the initial submit has a parent submit. There are two ways that rooted creations come about. One possibility is that a creator starts a new branch and chooses a parent submit as a root. Anyone may do that for any root branch at any time. The config can then either be chosen anew or inherited. Another possibility involves an ousted sprout, namely one that has been attempting to provide the next stable head in a lignification process. If that ousted sprout receives another submit, it turns into a proper branch. This branch is rooted in the branch for which it was a sprout. It inherits the branch config, but not the token and the branch contributors are the creator of the sprout and the content contributors of the branch that has been attempted to be merged. Any of those contributors can create submits to that ousted sprout and with that submit it turns into a proper branch where the parent entry is set during that conversion. Thus this mechanism for a rooted branch creation is indirect and can only be executed by the respective content contributors.

The creation of branches is permissionless. It is therefore a potential vector for a denial of service attack. The attacker can create a lot of branches and bombard other nodes with branch creation requests. One may leverage the token entry in a newly created branch to mitigate this risk. The attachment of a proof of token transfer in the token entry of the branch can function as a filter for sincere branch creation broadcasts.  

\subsubsection*{Merge}
In Lakat a merge is the inclusion of changes from one branch into another. There is a strict directionality in a Lakat-merge. Git distinguishes between this and other and in Lakat this corresponds to core and belt, where core is the pulling branch. Merging into a proper branch can only occur after a pull-request (see Proof-of-Review in Subsection \ref{ssc:por}). Twigs on the other hand can pull other branches using an approval of a fraction of its content contributors. The fraction is specified in the branch config. After a merge the belt branch may become stale. A stale branch cannot receive submits. Whether a branch becomes stale after a merge depends on the branch config (see Subsection \ref{ssc:configchange}).
A merge requires a merge submit (see Subsection \ref{ssc:submit}) and a cryptographic validation of the branch that is merged. When the conditions for a merge are not met, the merge submit cannot pass the cryptographic validation. For instance if the config of the pulling branch only allows conflictless submits and the belt branch has conflicted submits, then the merge is invalid.

In order to discuss which data buckets are included in the merge submit we briefly introduce the set theoretic slang of $A$ \textit{minus} $B$ for the set of elements in $A$ that are not in $B$. The set of elements that are in $A$ and $B$ is called \textit{intersection} of $A$ and $B$ and the set of elements that are in $A$ or $B$ is called the \textit{union} of $A$ and $B$. The respective notations are $A-B$, $A\cap B$ and $A\cup B$. We denote the set of submits of a branch $\mathfrak B$ by $\mathcal S_{\mathfrak B}$. We denote the set of data buckets in the data root of a submit $s$ by $\mathcal{B}_s$. Thus the set of data buckets in a branch $\mathfrak B$ with stable head $head(\mathfrak B)$ is $\mathcal{B}_{head(\mathfrak B)}$. 

We have already discussed that there is no many-to-one relation between buckets and branches (c.f. Figure \ref{fig:branchbucketrel}). There may be data buckets in core $\mathfrak C$ that are not in belt $\mathfrak P$ and there sure are data buckets in belt that are not in core, i.e. $\mathcal S_{\mathfrak P}-\mathcal S_{\mathfrak C}\neq \emptyset$ is not empty. 
One question that arises in the context of merges is how to combine disparate bucket sets and how to handle that on the level of submits. There are two possible design choices. Either all the submits of belt become submits of core and, consequently, also the buckets in $\mathcal S_{\mathfrak P}-\mathcal S_{\mathfrak C}$. Alternatively they stay submits of belt and the beforementioned buckets are included in the merge-submit's Merkle hash of the data trie. In the first scenario one is faced with the problem that the submits of belt all have immutable timestamps and parents. Rebasing those would require loosening those immutability conditions. In the latter scenario one needs to point to those submits from which data was included. It suffices to point to the belt's last submit before the merge. We opt for the second scenario. Unlike the first scenario, the second scenario has the peculiar situation that belt may have data buckets in common with core even though they do not share any submits, i.e. $\mathcal S_{\mathfrak P}\cap\mathcal S_{\mathfrak C}= \emptyset$ yet $\mathcal B_{\mathfrak P}\cap\mathcal B_{\mathfrak C}\neq \emptyset$. The only way this can happen in Lakat is if core and belt have pulled from the same branch or from branches that have a common submit in their histories \footnote{Here we make the distinction between the history of a branch and the set of submits of a branch. A branch may be rooted in another branch, but its history can go beyond the root.}.

%% file: integrationadaptability.tex
\section{Integration and Adaptability}
\label{sc:integrationadaptability}

\subsection{Onramping}
\label{ssc:onramping}

One of the objectives of Lakat is to transition academic publishing from a paper-formatted system to a cryptographically secure, collaborative and pluralistic system that allows for the continuous integration of research output. In order to achieve this objective, we believe that a transition should be as seamless as possible. The publishing system with isolated paper-formatted publications and intransparent review processes is an edge case of Lakat, an unsustainable and hacky one yet sufficient for onramping. We describe in which sense this is the case and how a transition could be achieved.

We can imagine a scenario for Lakat with a set of isolated branches. Each branch is controlled by a single legal entity, namely an academic journal. The academic journal is the content contributor, the storage contributor and the token contributor all in one. When a hypothetical researcher, say Alice (AI or human), wants to publish a paper, she has to send it to the journal. The branch that the journal  controls is simply the indexed collection of articles that have been submitted, respectively chained together cryptographically. For a journal to transition its content to such a branch state is anywhere between immediate -- by pointing the head of the branch to the storage locations of all content -- or a matter of running a script that creates a submit for each accepted article retrospectively.  Each paper is stored on a journal-controlled server, thus making the journal the sole storage provider. By adding the submission to the journal branch, the journal becomes the sole content contributor and retains all the rights of the contribution. The contribution is no longer owned by Alice at all. In this hypothetical oligarchic aberration of Lakat, a contribution is a submit with a single data bucket containing the paper. In summary, there is a way to map the classical publishing system into Lakat. Depending on the openness and licensing it might be difficult to either access or modify the content, but at least there is an entry point for the conversion.
Why is this unsustainable?  Given the design of Lakat, this branch would quite naturally undergo diversification through forking. At some point a researcher may create a branch rooted in that journal branch, which is but a click. Maybe the incentive structure provided by the journal is so strong that authors are willing to transfer all the rights to the journal voluntarily, but given Lakat's inherent ease of branching it will be a matter of time until a diversification is to be expected.

\subsection{Interfaces}
\label{ssc:interfaces}

We envision Lakat as a base layer for an open, pluralistic and collaborative publication system that progresses through continuous integration. As a base layer we strive to rely only on a bare minimum of other software and aim to have an interface for existing software or protocols. Here we provide an overview of the protocols and software that we plan to build upon or interface with.

We would like to use the libp2p library to implement Lakats demands for networking. Libp2p is a modular peer-to-peer networking stack which amongst others also contains Kademlia as a distributed hash table protocol. In particularly we would like to build a first client using the Rust implementation of libp2p. 

We would like to interface with mediawiki. Mediawiki is an evolving database schema with a php frontend that allows for the creation of knowledge databases such as wikipedia. There are various ways how Lakat could interface with mediawiki. The weakest form requires the conversion of the data contributions in Lakat to database entries in mediawiki. A stronger form converts also contributions to mediawiki into Lakat contributions. 

Regarding storage we aim to stay agnostic and leave the storage protocol as a configurable option. As options we consider IPFS, Ceramic (which is built on top of IPFS and anchored in Ethereum) or Urbit (lineDB). Regarding the token layer we aim to be agnostic here as well. Since this is an optional feature it is left to the branch contributors to decide and merge updates on their token transactions into their branch. We do recommend deploying tokens on the Polygon Layer 2 network though and plan to integrate this into the pipeline.

Regarding version control we would like to reduce the complexity of branch operations to a bare minimum in order to avoid security threats and reliance on other protocols. For the local consensus mechanism we believe that a heavily reduced set of operations is favorable. Nevertheless we would like to interface with existing version control systems such as git or radicle. We would like to interface with them in order to allow for the conversion of git or radicle branches into Lakat branches.

We would also like to allow for new branches to be turned into parathreads in Polkadot \cite{wood2016polkadot}. This would allow for the integration of Lakat into the Polkadot ecosystem. To this end one would need to create a pipeline to spin up a new parathread using Substrate together with a custom consensus protocol, namely the Lakat protocol. One would have to set the \textit{BlockImport} to a custom way of importing new submits into the key-value database. \textit{SelectChain} handles the finalization mechanism and would need to be set to the Lignification mechanism (See Subsection \ref{ssc:lignification}). One would also need to set the proof-of-review mechanism (See Subsection \ref{ssc:por}) in the \textit{Environment} option of the Substrate runtime.

%% file: conclusion.tex
\section{Conclusion}
\label{sc:conclusion}

The contributions presented in this paper are threefold. First we propose a process- and conflict-oriented academic publishing system that is based on a peer-to-peer network architecture and supports continuous integration across multiple branches. Second, we propose a new consensus mechanism for branched, permissionless systems, called Proof of Review. Third, we propose a new finality gadget, called Lignification, that is specifically designed for branched, permissionless systems. 

Regarding the first contribution of an architecture for an academic publishing system we provided a list of high-level requirements for such a system. We then argued that paper-based publishing has major shortcomings including the incentivization to withhold preliminary results, the tendency to wrap minor changes into an entire research paper, the lack of representation of the research process in the output, the creation of isolated content islands, the difficulty of tracking contributions and the barring of potential contributors. Based on that we proposed an architecture that meets the requirements and addresses these shortcomings. It is composed of a cryptographically linked data structure that is kept in a distributed key-value database, the entries of which are stored, updated and communicated using a peer-to-peer network protocol. Storage is partly outsourced to other protocols. The central data objects are branches and data buckets. All research content as well as updates and context information is stored in buckets. Branches on the other hand are cryptographically linked chains of changes to this underlying content. These changes are bundled up into submits, which are analogous to Git-commits. Every branch has its own staging area, where any type of branch interactions are waiting to be included. We distinguished three types of branches: proper branches which behave like production branches, twigs which are used like feature branches and sprouts which are temporary auxiliary branches. All branches are controlled by branch contributors. We distinguished four types of branch contributors: Content contributors who can prove to have submitted content to a given branch, review contributors who can prove to have pushed reviews to the branch, token contributors who can prove to have deposited funds into the branch and storage contributors who can prove to store data of that branch. Branch modifications happen via submits through a local consensus and finality mechanism amongst the branch contributors. In principle the Lakat specification does not require a particular choice of these mechanisms.

We put forth a particular local consensus mechanism for proposing changes in branched, permissionless systems. This constitutes the second contribution. The mechanism distinguishes between feature and production branches. Using Lakat terminology these are twigs and proper branches. Amendments to both types of branches are made through submits. Twigs ought to be used for small changes and quick iterations. Adding a submit requires a majority vote amongst the branch contributors, which in the case of a single contributor leads to no consensus rule at all. Proper branches on the other hand are used for larger changes. Amendments to proper branches involve merge submits, but the process is more akin to adding a block to a blockchain. First a pull-request is sent by a requesting branch to the staging area of the target branch. Any content contributor from the target branch may send a review commitment to the requesting branch and can participate in the review process through its role as a review contributor. Once a request has been reviewed and approved by a set number of reviewers a formally valid merge submit can be formed. Any contributor of the target branch can propose the amendment of the merge-submit to the proper branch. We called this process the Proof-of-Review (PoR).

Finally we proposed a mechanism to deterministically decide which proposed merge submits are eventually amended to the target branch. We put forth a finality gadget called "Lignification", which we presented as our third contribution. In order to decide which of the proposed merge submits become the head of the proper branch, we proposed a process using temporary proto-branches called sprouts. Every merge submit bears in it the possibility of becoming either the head of the proper branch or the head of forked branch. That is why we proposed to wrap the proposed merge submits into sprouts. There could be multiple proposed merge submits. In the lignification method there is a possibility for the branch contributors to issue vetoes and subsequently votes for case of multiple contesting merge submits. The lignification times and engagement times are respectively reserved for vetoing and voting. A broadcasting buffer is used to allow for network latency. 

We also mentioned Lakat's potential for interfacing with existing software and protocols, such as mediawiki, IPFS, Ceramic, Urbit, git, radicle, and Polkadot. Moreover, we drew a possible path to onramp users such as journals and scientists to Lakat. This further enhances its adaptability and integration capabilities, which is crucial for the system's adoption and growth, ensuring that it can evolve alongside technological advancements and changing academic needs.

Having presented the high-level concepts of a new publishing architecture, we are yet to find the concrete specifications and to test and study the elements in interaction. In a next step we are inviting anyone to collaboratively build client software for Lakat. We are currently in the process of building a Rust-client around the rust-libp2p library, called Lakat-OS. Please visit our \href{https://github.com/Lakat-OS}{Lakat Github Repository}.

%% file: acknowledgements.tex
\begin{acknowledgements}

I would like to thank Mahdi Kourehpaz for discussions and being a great critical mind and I would like to thank Andrei Taranu for discussions, support and for poking the idea here and there. Furthermore I appreciated the opportunity to present this idea at a workshop organized by the Basic Research Community in Physics in Spetember 2022, which provided a stimulating environment for discussion and feedback. In particular I am grateful for discussions with Benjamin Bose, Louis Garbe, Bernadette Lessel and Markus Penz. Finally I would also like to thank the Urbiter $\sim$dachus-tiprel for very insightful discussions in San Salvador and for the Urbit community to provide a great environment for discussions.

\end{acknowledgements}